\documentclass[12pt]{article}

\pdfoutput=1

\usepackage{color}
\usepackage{graphicx}

\usepackage{amssymb}
\usepackage{amsmath}

\newcommand{\rf}[1]{(\ref{#1})}
\newcommand{\beq}{\begin{equation}}
\newcommand{\eeq}{\end{equation}}
\newcommand{\bea}{\begin{eqnarray}}
\newcommand{\eea}{\end{eqnarray}}

\newcommand{\g}{\gamma}

\newcommand{\lam}{\lambda}

\newcommand{\bt}{\beta}
\renewcommand{\a}{\alpha}

\newcommand{\ep}{\varepsilon}

\newcommand{\om}{\omega}
\newcommand{\del}{\delta}
\newcommand{\Del}{\Delta}

\newcommand{\kp}{\kappa}

\newcommand{\ra}{\rangle}
\newcommand{\la}{\langle}

\newcommand{\pl}{\!+\!}

\newcommand{\ta}{{\tilde{\alpha}}}

\newcommand{\tkp}{\tilde{\kp}}

\begin{document}

\begin{center}
\vspace{24pt}
{\Large \bf Renormalization Group Flow in\\ \vspace{10pt} CDT}
\vspace{30pt}

{\sl J. Ambj\o rn}$\,^{a,b}$,
{\sl A. G\"{o}rlich}$\,^{a,c}$,
{\sl J. Jurkiewicz}$\,^c$,
{\sl A. Kreienbuehl}$\,^{b,d}$
and 
{\sl R. Loll}$\,^{b,e}$

\vspace{24pt}

{\footnotesize

$^a$~The Niels Bohr Institute, Copenhagen University\\
Blegdamsvej 17, DK-2100 Copenhagen \O , Denmark.\\
email: ambjorn@nbi.dk, goerlich@nbi.dk\\

\vspace{10pt}

$^b$~Radboud University Nijmegen\\
Institute for Mathematics, Astrophysics and Particle Physics (IMAPP),\\  
Heyendaalseweg 135, 6525 AJ Nijmegen, The Netherlands.\\
email: r.loll@science.ru.nl 

\vspace{10pt}

$^c$~Institute of Physics, Jagiellonian University,\\
Reymonta 4, PL 30-059 Krakow, Poland.\\
email: jerzy.jurkiewicz@uj.edu.pl\\

\vspace{10pt}

$^d$~Institute of Computational Science,
University of Lugano,\\
Via Giuseppe Buffi 13, CH-6900 Lugano, Switzerland.\\
email: andreas.kreienbuehl@usi.ch\\

\vspace{10pt}

$^e$~Perimeter Institute for Theoretical Physics,\\
31 Caroline St N, Waterloo, Ontario N2L 2Y5, Canada.\\
{email: rloll@perimeterinstitute.ca}
}

\vspace{24pt}
\end{center}


\begin{center}
{\bf Abstract}
\end{center}

\noindent 
We perform a first investigation of the coupling constant flow of the 
nonperturbative lattice model of four-dimensional quantum gravity 
given in terms of Causal Dynamical Triangulations (CDT).
After explaining how standard concepts of lattice field theory 
can be adapted to the case of this background-independent theory, 
we define a notion of ``lines of constant physics" in coupling constant space
in terms of certain semiclassical properties of the dynamically 
generated quantum universe.
Determining flow lines with the help of Monte Carlo simulations,
we find that the second-order phase transition line present in this theory
can be interpreted as a UV phase transition line if we allow 
for an anisotropic scaling of space and time.

\newpage

\section{Introduction}\label{introduction}

The formalism of Causal Dynamical Triangulations (CDT) provides a regularization
of the putative theory of quantum gravity \cite{physrep,cdtreviews}. 
Its underlying assumption is that the fundamental theory of quantum gravity 
can be understood purely in terms of quantum field-theoretical 
concepts. CDT quantum gravity shares this assumption with 
the asymptotic safety program, originally put forward by Weinberg \cite{weinberg}, which
was subsequently studied in a $(2\pl \ep$)-dimensional expansion \cite{kawai} and 
more recently with the help of the functional renormalization group equation \cite{FRG}.
Similarly, a key idea behind Ho\v{r}ava-Lifshitz gravity (HLG) \cite{horava} is to use
ordinary quantum field theory to construct quantum gravity, but to circumvent the usual
problem of non-renormalizability by explicitly breaking the four-dimensional diffeomorphism
invariance of the continuum theory with the introduction of a preferred time foliation. 
In this setting one can naturally introduce terms with higher {\it spatial} derivatives in the action
to render the theory renormalizable while keeping the theory unitary.  

Their common field-theoretic basis, as well as coinciding results on the spectral dimension of
spacetime on Planckian scales \cite{spectral} and a similar phase structure of CDT and HLG \cite{phase,phase1}
make it natural to try to relate the three approaches --
causal dynamical triangulations, asymptotic safety and Ho\v{r}ava-Lifshitz gravity --
more directly.\footnote{More distant relatives of CDT are
group field theory \cite{GFT} and so-called tensor models
\cite{tensor}, which in specific limits can generate 
triangulations. These models are presumably more closely related to (Euclidean) Dynamical 
Triangulations \cite{aj,am} than to CDT.} Interesting examples of this include the formulation of a functional renormalization
group equation for foliated spacetimes \cite{hl-rg} and its application to projectable HLG at low energies \cite{crs}, 
and an extension of CDT quantum gravity by
the explicit addition of higher spatial derivative terms (albeit at this stage only in
three spacetime dimensions \cite{carlip}). Note that HLG does not appeal to an asymptotic safety
scenario for the theory to make sense at high energies.

Although the distinguished notion of proper time of CDT looks superficially similar to the time foliation 
in HLG, its status is different because CDT does not possess any residual diffeomorphism invariance, 
which therefore cannot be broken either. The role of time in CDT was recently clarified further in a study in three
dimensions, where it was verified explicitly that key results of CDT quantum gravity continue to hold in a version of the
theory which does not possess preferred simplicial hypermanifolds that can be identified with surfaces of
constant time \cite{jordanloll}. This provides strong evidence that the notion of proper time that is
naturally available in standard CDT is simply a convenient parameter to
(partially) describe the spacetime geometry, and that its presence does not skew the results of the theory in an
unwanted way. Of course, also this ``non-foliated" version of CDT incorporates microscopic causality conditions, 
implying an asymmetry between time and spatial directions that persists after Wick-rotating, just like in standard CDT. 
It is therefore conceivable that in part of the coupling constant space 
\cite{phase} the nonperturbative effective quantum action of CDT 
can be related to an anisotropic action of HLG-type, even though in the former no higher-order spatial derivative terms are added 
explicitly to the bare classical Einstein-Hilbert action. 
Let us also point out that the built-in unitarity of the CDT formalism -- resulting from 
a well-defined transfer matrix \cite{original} -- 
is likely to affect the functional form of the dynamically generated quantum action, in a way we currently do not
control explicitly.

In this article, we present a first attempt at establishing a concrete renormalization
group flow in four-dimensional CDT quantum gravity
(in the standard version and without higher-derivative terms in the bare action), assuming a
straightforward identification of lattice proper distances with continuum proper distances.
More specifically, with the help of computer simulations
we determine trajectories of constant physics -- interpreted in a specific way in terms of
semiclassical observables we have at our disposal -- in the 
coupling constant space
spanned by the bare coupling constants of the lattice theory. Moving along these lines in the direction of smaller
lattice spacing, we do not find evidence that they run into the second-order phase transition line, with the
possible exception of the triple point of the phase diagram, where three transition lines meet. 
A slightly more general ansatz that allows for a relative scaling of time and space as the second-order
transition is approached leads to a more interesting result, which can be interpreted as a proper UV limit. 
-- In terms of procedure and first results, our investigation provides a reference frame and opens the door to
a further systematic study of renormalization group flows in CDT and perhaps other nonperturbative
lattice formulations of quantum gravity. 
This will involve more sophisticated arguments for an appropriate relative scaling of
time and space near the phase transition, and hopefully a wider range of observables to provide alternative
definitions of what it means to ``keep physics constant".

\section{Causal Dynamical Triangulations}

CDT is a theory of fluctuating geometries, which at the regularized level are represented by
triangulated, piecewise flat spacetimes. 
It can be viewed as a lattice theory in the sense that the length assignments to the one-dimensional
edges (links) of a given triangulation completely determine the piecewise flat geometry.\footnote{Let us emphasize that
these geometries are perfectly continuous (and not discrete, as is sometimes stated), 
despite the fact that curvature is distributed on them 
in a singular manner.}
As already mentioned, a well-behaved causal structure is implemented on each Lorentzian triangulation
with the help of a global time foliation that is distinguished in terms of the simplicial structure. 
One sums over these geometries in the path integral, where the action is given by 
the Einstein-Hilbert action in Regge form, suitable for piecewise linear geometries (see
the review \cite{physrep} or the original articles \cite{original} for further details). 
All triangulations can be obtained by suitably gluing together
two types of building blocks, the so-called (4,1) and (3,2) four-simplices, leading  (after Wick rotation) to
a very simple form for the Euclidean Regge action $S_E$, namely,
\begin{equation}
S_E=-(\kappa_0+6\Delta) N_0+\kappa_4 (N_{4}^{(4,1)}+N_{4}^{(3,2)})+
\Delta (2 N_{4}^{(4,1)}+N_{4}^{(3,2)}),
\label{1.1}
\end{equation} 
where $N_{4}^{(4,1)}$ and $N_{4}^{(3,2)}$ are the numbers of 
four-simplices of type (4,1) and (3,2) respectively, 
and $N_0$ is the number of vertices in the triangulation. The parameter $\kp_0$
is proportional to $a^2/G_0$ where $G_0$ is the bare gravitational
coupling constant and $a$ denotes the length of (spatial) links. 
Similarly, $\kp_4$ is proportional to the bare cosmological constant but will 
play no role here, since we will keep the number of four-simplices (almost)
constant during the Monte Carlo simulations of the CDT lattice system. 

The parameter $\Delta$ appearing in the action (\ref{1.1}) requires a more detailed discussion. 
There are two types of edges that occur in the Lorentzian-signature triangulations
before everything is Wick-rotated, spacelike links with squared length $a^2$
and timelike links with negative squared length $a^2_t \! =\! -\a a^2$, where the
parameter $\a >0$ quantifies the relative magnitude of the two.
We then perform a rotation to Euclidean signature 
by analytically continuing $\a$ in the lower-half complex plane from 
$\a$ to  $-\a = \tilde{\a}$, so that
\beq\label{1.2}
a_t^2 = -\a a^2 ~~~~\mapsto ~~~~~~a_t^2 = \tilde{\a} a^2,~~~~\tilde{\a} >0.
\eeq
The original, Lorentzian Einstein-Hilbert action in Regge form depends on $\a$ and 
satisfies $iS_L(\a)\! =\! -S_E(-\a)$ when rotating from Lorentzian 
to Euclidean signature. The Euclidean action $S_E$ is now 
a function of $\tilde{\a}$ (see \cite{physrep} for details). 
It can be parametrized in the form 
\rf{1.1}, where $\Del$ is now a function of $\tilde{\a}$, normalized such
that the case of uniform edge lengths, $\ta=1$, corresponds to $\Del=0$.

At this stage $\Del$ is not a coupling constant, but only
a parameter in the action. Even for $\Del$ different from zero the action continues to be
the Euclidean Regge-Einstein-Hilbert action, merely reflecting the 
fact that some links are assigned a different length. 
However, in the effective quantum action $\Del$ will appear as a
coupling constant. The reason why this can happen is that the choice of 
coupling constants for which interesting fluctuating geometries 
are observed is far from the semiclassical region. 
In this nonperturbative region the measure used in the 
path integral becomes as important as the classical action, 
and $\Del$ will effectively play the role
of a coupling constant. We refer again to \cite{physrep} for a detailed
discussion, and examples of nongravitational lattice models where one encounters 
a similar situation.  
In view of this, the coupling constant space of CDT quantum gravity is spanned by $\kp_0$ and $\Del$. 

\begin{figure}[t]
\vspace{-5cm}
\begin{center}
\includegraphics[width=1\textwidth]{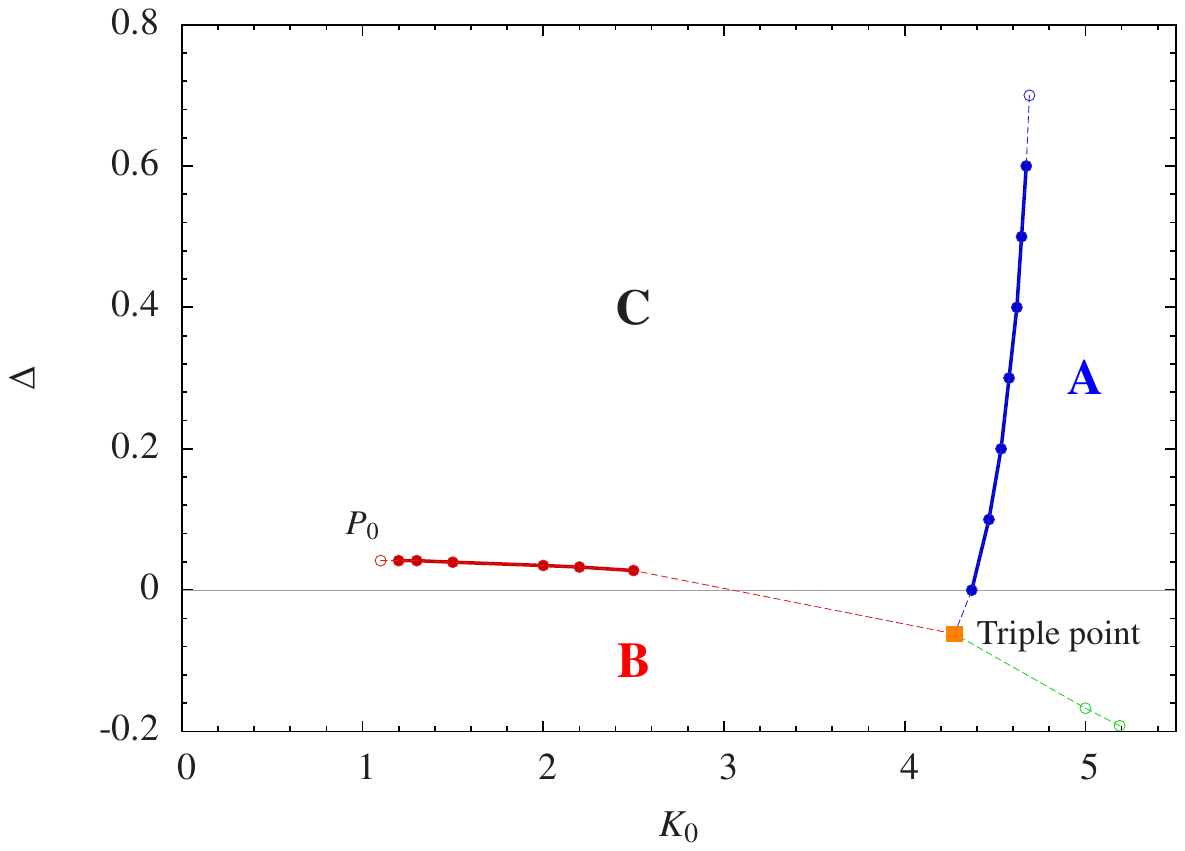}
\end{center}
\vspace{-10cm}
\caption{Phase diagram of CDT quantum gravity in four dimensions.}
\label{fig0}
\end{figure}

For reference, we are showing in Fig.\ \ref{fig0} the corresponding phase diagram, already reported in \cite{ajl,phase1}. 
It has three phases, denoted by $A$, $B$ and $C$. 
Previous studies have shown that only phase $C$ is interesting from the point of view of quantum gravity,
in the sense that only there one seems to find quantum fluctuating geometries which are macroscopically
four-dimensional. The properties of the quantum geometry in this phase have  
been studied in great detail \cite{ajl1,ajl,agjl,agjl-etal}.

In the present work, we will follow standard lattice procedure by trying to trace the flow 
of the bare coupling constants inside phase $C$
when we take the lattice spacing $a$ to zero, while keeping physics constant. 
We know from \cite{phase1} that the phase transition line 
separating phases $B$ and $C$ is of second order, while phases $A$ and $C$ are separated
by a first-order transition. Our expectation is therefore that the flow lines will approach
this second-order transition line when $a$ goes to zero and continuum 
physics is kept constant.

\section{Identifying paths of constant physics}\label{sec:constant}

For the purpose of illustration, consider a $\phi^4$-lattice scalar field theory with bare (dimensionless) 
mass term $m_0$ and bare dimensionless 
$\phi^4$-coupling constant $\lam_0$. Correspondingly, the effective action has 
a renormalized mass $m_{R}$ and a renormalized coupling constant $\lam_R$.  
Let us assume that $\lam_R$ is defined according to some 
specific prescription in terms of the four-point function. Similarly, 
assume that $m_R$ is defined by some prescription related to the 
two-point function, for example the exponential fall-off 
of the connected two-point function\footnote{As usual in a lattice set-up, there is 
the question of lattice artifacts when defining $m_R$ and $\lam_R$, 
due to the finiteness of the lattice spacing and accompanying
discretization effects.
In the discussion below we ignore such technical issues 
because our focus will be on the essence of the
renormalization group flow of the bare lattice coupling constants.}. 
One can thus write $m_R a\! =\! 1/\xi$, 
where $\xi$ is the correlation length measured in lattice units $a$. 
This relation specifies how one should scale the lattice spacing 
$a$ to zero as a function of the correlation length $\xi$ in order for 
$m_R$ to stay constant. Once the actual value of $m_R$ has been supplied
from the outside, say by comparison with experiment, the value of 
$a(\xi)$ is fixed in physical units by measuring $\xi$. 

In order to define a continuum limit where $a(\xi) \to 0$ while $m_R$ is kept fixed 
one needs a divergent correlation length $\xi$, in other words, a phase transition point or 
phase transition line 
of second order in the $(m_0,\lam_0)$-coupling constant space. 
The lattice $\phi^4$-theory has 
such a second-order phase transition line. Choosing specific initial values 
$m_0(0)$ and $\lam_0(0)$ for the bare coupling constants, 
performing the functional lattice integral
will determine the renormalized coupling 
$\lam_R\! =\! \lam_R(m_0(0),\lam_0(0))$ corresponding to these values. 
The requirement that $\lam_R(m_0,\lam_0)$ stay constant when changing $m_0$ and 
$\lam_0$ then defines a curve $(m_0(s),\lam_0(s))$ in the plane spanned by the
bare coupling constants, where $s$ is an arbitrary curve parameter. 

Along this curve the correlation length $\xi$ will change. Assuming for simplicity
that $\xi$ is a monotonic function of $s$, one can parametrize the curve by $\xi$ 
instead. Moving along the curve in the 
direction of increasing $\xi$ will in general lead to
the second-order phase transition line where $\xi$ becomes infinite.
At the same time, because of $a(\xi)\! =\! 1/(m_R \xi)$,
the UV cut-off $a$ will decrease. 
If the curve reaches the transition line at a point $\lam^*_0$, 
this point will be a UV fixed point for the $\phi^4$-theory, corresponding to a 
renormalized mass $m_R$ and a renormalized coupling constant 
$\lam_R$, since approaching it one has $a(\xi) \to 0$. 
However, it can happen that a curve of constant $\lam_R$ does not reach
the second-order phase transition line. If one cannot find a single
curve of constant $\lam_R$, for any starting point $(m_0,\lam_0)$, which reaches such a critical point,
one would conclude that the theory does not have a UV completion with a finite value 
of the renormalized coupling constant $\lam_R$.
For the four-dimensional scalar $\phi^4$-theory this turns out to be the case.

Assume for the sake of the argument that there {\it is} a UV fixed point $\lam_0^*$ 
somewhere on the 
second-order phase transition line\footnote{Note that in formulas \rf{2.1}
and \rf{2.2} below it is assumed that $\lam^*_0 \neq 0$. If $\lam^*_0 =0$ the 
fixed point is Gaussian and the formulas have to be modified appropriately.}. 
The $\bt$-function then has a zero there, $\bt(\lam_0^*)\! =\! 0$,
since at fixed $m_R$ and $\lam_R$ the coupling $\lam_0(\xi)$ stops running for 
$\xi \to \infty$. 
Approaching the fixed point along such a trajectory, 
the behaviour of $\lambda_0$ near $\lam_0^*$ is described by 
\beq\label{2.1}
\lam_0(\xi) = \lam_0^* + const. \;\xi^{\bt'},~~~
\bt'= \frac{d \bt}{d \lam_0}\Big|_{\lam_0=\lam_0^*}.
\eeq
In the CDT quantum gravity theory it will be convenient to 
analyze the flow of bare coupling constants for fixed 
continuum physics under the additional assumption that 
the {\it physical} volume of spacetime is fixed and finite.
With this in mind, one can reformulate the above coupling constant flow
in ordinary lattice field theory in terms of so-called finite-size scaling.  
Consider the case of $d$ dimensions
and introduce a dimensionful physical $d$-volume $V_d$ by $V_d\! :=\! N_d a^d$,
where $N_d$ is the total number of $d$-dimensional elementary building blocks
(hypercubes on a cubic lattice, simplices on a triangular lattice). 
We want to make sure that $V_d$ can be viewed as constant along 
a trajectory of the kind described above, with $m_R$ and $\lam_R$
kept fixed, in the continuum limit as $a(\xi) \to 0$. 
This can be achieved by keeping  
the ratio between the linear size $L\! =\! N_d^{1/d}$ of the lattice 
``universe'' and the correlation length $\xi$ fixed. In terms of
the renormalized mass $m_R$ and the lattice spacing $a(\xi)$ the ratio 
can also be written as 
\beq\label{jnew1}
\frac{\xi^d}{N_d}= \frac{ 1}{(a(\xi)m_R)^d N_d} = \frac{1}{m_R^d V_d}.
\eeq 
Accordingly, moving along a trajectory of constant $m_R$ and  $\lam_R$ in the 
bare $(m_0,\lam_0)$-coupling constant plane and changing $N_d$ 
ensures that the quantum field theory in question has 
a finite continuum spacetime volume $V_d$. Furthermore, the equality \rf{jnew1} implies 
that the dependence on the correlation length $\xi$ in \rf{2.1} can be substituted by a dependence on
the linear size $N^{1/d}$ in lattice units of the spacetime, leading to
\beq\label{2.2}
\lam_0(N_d) = \lam_0^* + const. \;N_d^{\bt'/d}.  
\eeq
We noted above that the absence of a UV fixed point is signaled by the fact that
no curve of constant $\lam_R$ reaches the phase transition line. In this case
the correlation length $\xi$ along curves will not go to infinity 
and the lattice spacing will not go to zero. Restated
in terms of the discrete lattice volume it means that $N_d$ will not go to infinity. 

We have outlined in this section in some detail how to define and follow lines of
constant physics in the $\phi^4$-lattice scalar field theory, because we want to
apply the same technique to understand the UV behaviour of the lattice quantum gravity
theory. Of course, it should be emphasized that the two theories differ in important ways.
First, because $\phi^4$-theory is renormalizable in four dimensions, we know
a priori that it suffices to study the flow in the bare couplings $m_0$ and $\lambda_0$:
if no UV fixed point is found along lines of constant $\lambda_R$ in the $(m_0,\lambda_0)$-plane, 
it does not exist.
On the other hand, gravity is not renormalizable, and restricting the search for a UV fixed point
to the two-dimensional coupling constant 
space spanned by $(\kappa_0,\Delta)$ -- although 
suggestive because of the observed second-order transition line --
may ultimately not be sufficient.

Second, while the meaning of lines of constant physics is relatively straightforward in $\phi^4$-theory,
the same cannot be said about this concept in nonperturbative and background-independent
quantum gravity, because any measure of 
length one uses is defined in terms of geometry, which is subject to the dynamics of the theory. 
As will become clear in the remainder of this paper, defining lines of constant physics in terms of 
suitable geometric observables needs
considerable care and is at this stage much more tentative than in the case of scalar field theory.

\section{Application to nonperturbative gravity}\label{sec:apply}

In the present application to quantum gravity, we will 
use the coupling constant flow in the form \rf{2.2}, staying at 
a constant spacetime volume $V_4 \! =\! N_4 a^4$ for the universe, where
$N_4$ is the number of four-simplices\footnote{Strictly speaking, we are keeping the number
$N_4^{(4,1)}$ of four-simplices of type $(4,1)$ constant, see \cite{agjl} for a discussion. The distinction is
not important for our present analysis.}. 
How can we make sure that it is consistent to view $V_4$ as constant when we 
increase the lattice volume $N_4$? 
In the case of ordinary field theory we achieved this by using the physical correlation length 
as a fixed yardstick and requiring $m_{R}^d V_d$ to remain constant.
Since in the CDT pure gravity model we do not have a similar simple correlation length at our disposal,
we need to find another indicator of constant physics.

In phase $C$, at least somewhat away from the $B$-$C$ phase 
boundary, the three-volume profile of the universe is to excellent
approximation given by \cite{agjl}
\beq\label{2.3} 
\la N_3(i) \ra_{N_4} = N_4\; \frac{3}{4}\; \frac{1}{\om N_4^{1/4}} \; 
\cos^3\left( \frac{i}{\om N_4^{1/4}}\right), ~~~~
|i| \leq \frac{\pi}{2} \om \, N^{1/4}_4,
\eeq
and the variance of the spatial volume fluctuations $\del N_3(i)\! :=\! N_3(i)-\la N_3(i) \ra$ by
\beq\label{2.4}
 \la (\del N_3(i))^2 \ra_{N_4} = 
\g^2 \; N_4 \; F\left( \frac{i}{\om N_4^{1/4}}\right),
\eeq
for a specific function $F$, whose details are not important for the discussion at hand.
Both profiles are functions of the lattice time $i$. The number of spacelike three-simplices at fixed integer time
$i$ is denoted by $N_3(i)$, and the parameters $\omega$ and $\gamma$ depend on the geometric properties
of the triangular building blocks and the bare coupling constants $\kp_0$ and $\Del$.

The  profiles \rf{2.3} and \rf{2.4} represent 
finite-size scaling relations, and show in the first place that the time extension of the universe
scales like $N_4^{1/4}$ and its spatial volume at a given time like $N_4^{3/4}$, as one would 
expect from a four-dimensional spacetime. 
This might seem like a triviality since we started out with four-dimensional building blocks, but
in a set-up where no background geometry is put in by hand it 
is not: all our results are extrapolated to an infinite limit ($N_4\to \infty$), and in this limit 
nonperturbative contributions from the summed-over path integral histories play an
important role in bringing about the final outcome. To illustrate the point, no four-dimensional macroscopic
scaling behaviour is found in phases $A$ and $B$ of the present model, although 
they are of course based on exactly the same (microscopically four-dimensional) building blocks.
Similarly, one may in principle find deviations from such a scaling inside phase $C$
when getting close to the second-order transition between phases $B$ and $C$.

The data \rf{2.3} and \rf{2.4} extracted from the Monte Carlo simulations in phase $C$ at fixed 
lattice volume $N_4$
allow us to interpret the ground state of geometry as a macroscopically four-dimensional 
quantum universe with a definite average volume profile and a definite behaviour of the 
average quantum fluctuations of the spatial volume around it. Moreover, making a specific
identification of continuum proper time with lattice proper time (by fixing a relative constant for
given values of the bare couplings),
these properties are characteristic for a de Sitter universe \cite{agjl}. 

Sufficiently far away from the phase
boundaries of phase $C$ the data summarized in relations \rf{2.3} and \rf{2.4} is compatible 
with the discretized action
\beq\label{2.4c}
S_{discr} = k_1 \sum_{i} \left(\frac{(N_3(i+1) -N_3(i))^2}{N_3(i)} +
\tilde{k} N_3^{1/3}(i)\right),
\eeq
which was reconstructed from measuring the correlation function of spatial three-volumes \cite{agjl}
and has the form
\beq\label{2.4e}
\left\la \delta N_3(i)\,
\delta N_3(i')
\right\ra_{N_4} =\gamma^2 
N_4 \; F \left( \frac{i}{\om N_4^{1/4}},\frac{i'}{\om N_4^{1/4}}\right),
\eeq
where it is understood that the function $F$ for identical arguments coincides with the function $F$
on the right-hand side of eq.\ \rf{2.4}.
For sufficiently large $N_4$ and to first approximation, the measured parameters $k_1$ and $\tilde{k}$ 
in the reconstructed action 
(\ref{2.4c}) were shown to be independent of $N_4$ and the 
coefficient $\g$ in \rf{2.4} was shown to be related to $k_1$ by 
\beq\label{2.4f}
\g \propto \frac{1}{\sqrt{k_1}}.
\eeq
Phrased differently, for appropriate choice of the coupling $\tilde{k}$ the 
classical solution to the discretized action \rf{2.4c}, solved
under the constraint of fixed $N_4$, is well approximated by 
the observed distribution $\la N_3(i) \ra_{N_4}$ of \rf{2.3}.
In addition, the observed behaviour of the volume fluctuations, eqs.\ \rf{2.4} and \rf{2.4e},  
is well described by expanding the action \rf{2.4c} to quadratic order
around the average profile \rf{2.3}, thus leading to \rf{2.4f}. Note finally that 
the coupling constant $\tilde{k}$ is a function 
of $\om$ if the distribution \rf{2.3} is to represent the 
local minimum of $S_{discr}$ for large, but fixed $N_4$, namely
\beq\label{2k}
\tilde{k}= 9 \left(\frac{3 }{4\om^4}\right)^{2/3}.
\eeq 

A natural starting point for trying to relate the above results to continuum physics is 
to compare the effective action \rf{2.4c} for the spatial three-volume
(constructed from numerical ``observations'') 
with a minisuperspace action for the scale factor of a homogeneous, isotropic
universe with spatial slices of the same $S^3$-topology.
We can then ask which 
continuum minisuperspace actions can be matched to an emergent background like \rf{2.3}.
The line element of (Euclidean) minisuperspace is 
\beq\label{2.4p}
ds^2 = N^2(t) dt^2 + a^2(t)d\Omega_3^2,
\eeq  
where $a(t)$ is the scale factor, $N(t)$ the lapse function and $d\Omega_3^2$ the 
line element on the unit three-sphere,
such that the spatial volume at time $t$ is $V_3(t) =2\pi^2 a^3(t)$.

As we have already argued in the introduction, Ho\v{r}ava-Lifshitz gravity provides a 
natural and potentially useful reference frame for nonperturbative properties of CDT quantum gravity.
Also in our present analysis of the renormalization group flow we will use an extended class
of reference metrics of type \rf{2.4p}, including minisuperspace models 
of Ho\v{r}ava-Lifshitz type. 

Recall that the quadratic part of the action of projectable\footnote{Because of the symmetry 
reduction to minisuperspace we are considering below, the difference between projectable and nonprojectable 
HLG will not play a role here.} HLG in four dimensions in terms of the three-metric $g_{ij}(x,t)$ and 
the extrinsic curvature $K_{ij}(x,t)$ reads
\beq\label{2.4q}
S_{cont} = {\tilde{\kp}} \int dt\, d^3 x N(t)\sqrt{g}\; 
( K_{ij} K^{ij} -\lam K^2  +
\tilde{\del}\, {}^{(3)}\! R ),
\eeq
where $N(t)$ is the lapse function and
${}^{(3)}\! R$ is the intrinsic scalar curvature of the spatial three-geometry. 
For the parameter values $\lam\! =\! 1$ and $\tilde{\del}\! =\! -1$ one 
obtains the standard form of the Euclidean Einstein action, in which case one
can identify $\tilde{\kp}\! =\! 1/(16\pi G)$, where $G$ is the 
gravitational coupling. The three terms in parentheses on the right-hand side of
\rf{2.4q} are separately
invariant under foliation-preserving diffeomorphisms, the invariance group of HLG.

Using the metric ansatz \rf{2.4p}, with $a(t)$ re-expressed in terms of
$V_3(t)$, the continuum HLG action \rf{2.4q} becomes
\beq\label{2.4s}
S_{cont} = {\kp} \int\, dt\, N(t)\; \Big( 
\frac{ {\dot{V_3}}^2}{N^2 V_3}+  \del\, V_3^{1/3} \Big), ~~~
\frac{\del}{\tilde{\del}} = \frac{18 (2\pi^2)^{2/3}}{1-3\lam},~~~
\frac{\kp}{\tilde{\kp}} = \frac{1-3\lam}{3}. 
\eeq
Firstly, the equation of motion derived for $V_3(t)$ from the action \rf{2.4s}, 
under the constraint that the total four-volume is $V_4$, is solved by 
\beq\label{2.4t}
V_3(\tau) = V_4\, \frac{3}{4} \,
\left(\frac{8\pi^2}{3\chi^3V_4}\right)^{1/4}
\cos^3 \left( \left(\frac{8\pi^2}{3\chi^3 V_4}\right)^{1/4} \, \tau\right),
~~~N =const.,
\eeq
which we have written in a form that facilitates comparison with the lattice
expression \rf{2.3}. It is of course precisely the match of the lattice results with 
a classical $\cos^3$-profile (\ref{2.4t}) that allows us to identify lattice time with a continuum
time $t$, which is a {\it constant} multiple of continuum proper time $\tau$,
\beq\label{2.4u1}
\tau = N \,t,\;\;\; N =const.
\eeq
The parameter $\chi$ in relation (\ref{2.4t}) is defined as
\beq\label{2.4u}
\chi^2 = \frac{9 (2\pi^2)^{2/3}}{ \del}.
\eeq

Computing the scale factor $a(t)$ corresponding to the volume profile 
(\ref{2.4t}) and substituting it into the line element \rf{2.4p} one obtains
\beq\label{2.4x}
ds^2= d\tau^2 + R^2 
\cos^2 \left( \frac{\tau}{\chi\, R}\right) \,d\Omega_3^2,~~~~
R = \left( \frac{3{V}_4}{8 \pi^2 \chi}\right)^{1/4}.
\eeq
Unless $\chi$ equals its general relativistic value $\chi=1$ this describes a deformed
four-sphere with time extension $\pi \chi R$ and
spatial extension $\pi R$, $R$ being the (maximal) radius of the spatial 
three-sphere\footnote{The geometry of the deformed four-sphere is 
not smooth at $\tau = \pm \chi\pi R/2$. The intrinsic curvature is 
discontinuous but integrable at these points.}. 

Next, comparing the continuum expressions \rf{2.4s}--\rf{2.4u} with the corresponding
lattice expressions \rf{2.3} and \rf{2.4c}
and assuming ${V}_4\propto N_4a^4$,
one is led to the identifications 
\beq\label{2.4v}
\tau_i \propto \left(\frac{\chi^{3/4}}{\om}\right) i\cdot a,~~~~~
k_1 \propto \left(\frac{\om}{\chi^{3/4}}\right)^2\, a^2\kp,
~~~~~~\tilde{k} \propto \left(\frac{\chi^{3/4}}{\om}\right)^{8/3} \del.
\eeq
We note that in the transition from lattice to continuum data only the ratio of $\omega$ and $\chi^{3/4}$ appears.
The first relation in \rf{2.4v} reiterates our earlier assertion that the continuum proper time can be viewed
as proportional to the integer lattice time multiplied by the lattice spacing, where the 
said ratio is now seen to enter.

Following the logic outlined at the beginning of this section, we would now like 
to define a path of constant continuum physics in the coupling constant space spanned
by $(\kp_0,\Del)$. In doing this, we want to keep the total four-volume $V_4 \propto N_4 a^4$ fixed. 
This will enable us to take the lattice spacing $a \to 0$ by changing $N_4$, 
a parameter we can control explicitly. Our definition of what constitutes ``constant physics"
will rely on the assumptions that (i) throughout phase $C$ the behaviour of the three-volume 
is described adequately by the (semi-)classical 
continuum formulas derived above, and (ii) we can associate space- and time-like lattice units with
continuum proper distances and proper times in a way that inside phase $C$ is independent of $\kappa_0$
and $\Delta$. More precisely, regarding this latter point it is sufficient to make the weaker assumption that the
{\it ratio} of unit proper distance and unit proper 
time is a fixed number times the speed of
light $c$ throughout coupling constant 
space.\footnote{Note that in the continuum expressions
used above we have set $c=1$, $\hbar=1$ everywhere. Re-introducing them makes it explicit that
the parameter $\chi$ defined in eq.\ (\ref{2.4u}) above has the dimension of an inverse velocity,
and that the product $c\cdot\chi$ is therefore dimensionless.} This is equivalent to keeping fixed the ratio 
$\omega/\chi^{3/4}$ in relations (\ref{2.4v}).

Under these assumptions, keeping $\omega$ constant in the simulations implies
a constant $\chi$ and thus a constant volume profile, giving us one criterion for constant,
macroscopic physics. However, keeping $\om(\kp_0,\Del)$ fixed is not sufficient to
ensure that the emergent continuum universe 
is unchanged in the limit $N_4 \to \infty$. Denoting the typical size of volume fluctuations
by $|\delta N_3(i)|\! :=\!\langle N_3(i)N_3(i)\rangle_{N_4}^{1/2}$ and analogously for $|\delta V_3(\tau)|$, one has
\beq\label{2.6}
\frac{|\del V_3(\tau_i)|}{V_3(\tau_i)} = \frac{|\del N_3(i)|}{\langle N_3(i)\rangle} \propto 
\frac{ \g(\kp_0,\Del)\, \om(\kp_0,\Del)}{N_4^{1/4}}~
\left(\propto \frac{ \chi^{3/4}}{\sqrt{\kp} \,V_4^{1/4}}\right),
\eeq
where the result in parentheses follows from relations \rf{2.4f} and 
\rf{2.4v}, and the scaling should be understood for fixed times $\tau$.
In view of the proportionality $\tau_i \propto i/N^{1/4}_4$ from \rf{2.4v} above, 
the discrete time label $i$ used 
in $N_3(i)$ and $\del N_3(i)$ should change proportional to $\tau N^{1/4}_4$
when changing $N_4$.   
According to our assumptions the three-volume profile $V_3(\tau)$ and the
fluctuation size $|\del V_3(\tau)|$ are physical quantities, and the ratio
$|\del N_3(i)|/N_3(i)$ (with the interpretation of $i$ just given)
must therefore remain constant along any path of constant physics in the space of bare 
coupling constants.

First, note that staying at a given point $(\kp_0,\Del )$ while taking 
$N_4\to \infty$ does {\it not} correspond to constant continuum physics.
Rather, according to (\ref{2.6}) it describes a 
situation where $V_3(\tau)$ (and $V_4$) go to infinity, 
and the fluctuations around this macroscopic universe 
become ever smaller relative to $V_3(\tau)$. 
Since we have already established that $\om(\kp_0,\Del)$ must be 
kept fixed along a
trajectory of constant physics, eq.\ \rf{2.6}
implies that as $N_4 \to \infty$ we must follow a path
$(\kp_0(N_4),\Del(N_4))$ satisfying  
\beq\label{2.7}
  \g(\kp_0(N_4),\Del(N_4)) \propto N_4^{1/4},  
~~~~\om(\kp_0(N_4),\Del(N_4)) = const.
\eeq
This pair of conditions can be regarded as the CDT 
equivalent of keeping $V_4$ constant 
in scalar field theory by insisting that the correlation length satisfies $\xi \propto N_d^{1/d}$, 
as discussed above. Furthermore, we read off from relation \rf{2.6} that 
the conditions \rf{2.7} are consistent with a physical situation where also 
the gravitational coupling constant $\kp$ is kept fixed. 
In the next section we will
investigate whether it is possible to satisfy \rf{2.7} in the 
limit as $N_4 \to \infty$.

\section{Measuring indicators of constant physics}

\begin{figure}[t]
\vspace{-1cm}
\begin{center}
\includegraphics[width=0.5\textwidth]{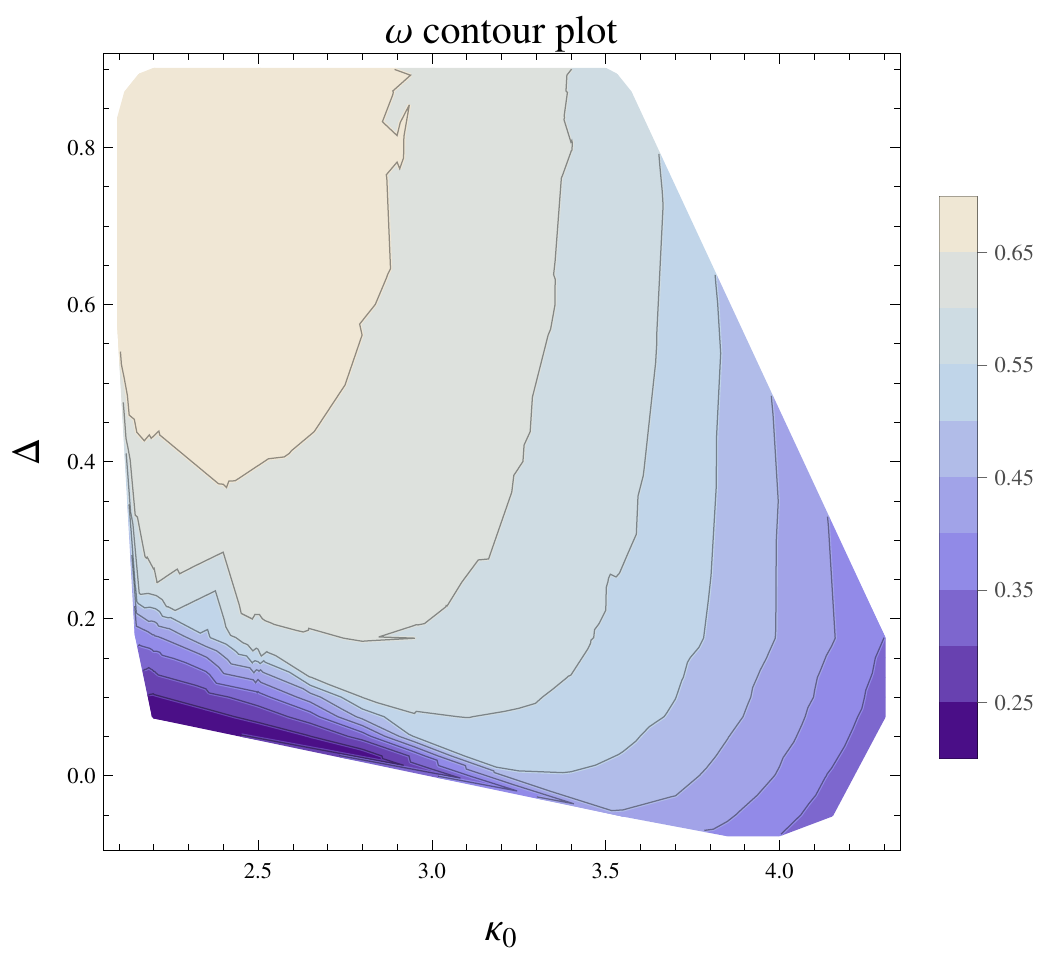}\includegraphics[width=0.5\textwidth]{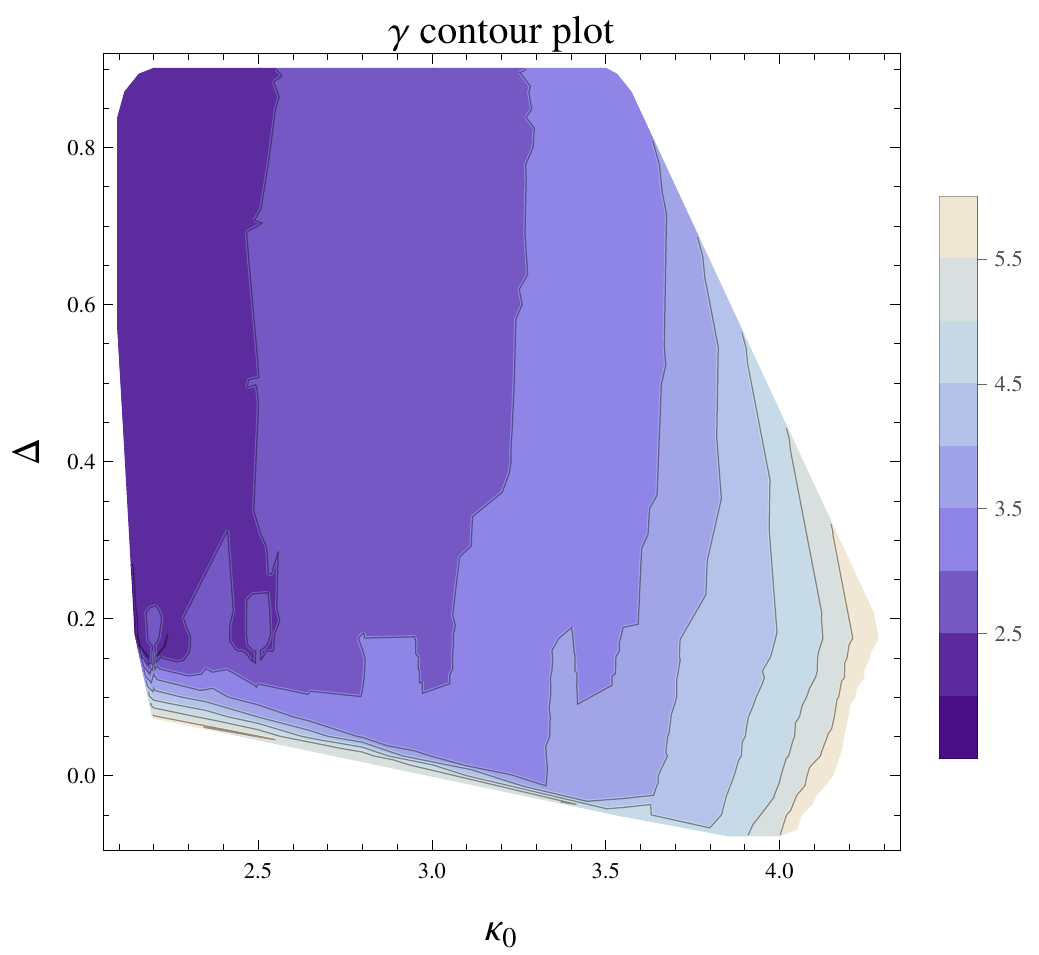}
\end{center}\caption{Contour plots in phase $C$ 
of the parameters $\om(\kp_0,\Del)$ (left) and  
$\g(\kp_0,\Del)$ (right), used to characterize trajectories of
constant physics.  They behave roughly oppositely, $\omega$
decreasing and $\g$ increasing toward the bottom right. Their product  
$\om\cdot\gamma$ changes only moderately, as illustrated by 
Fig.\ \ref{fig2} below.} 
\label{fig1}
\end{figure}

In phase $C$ of the CDT phase diagram we have performed a systematic study 
measuring the distributions $N_3(i)$ for a fixed number $N_4$ of building blocks. By fitting, following the 
procedure outlined in \cite{agjl}, we can determine $\om(\kp_0,\Del)$ and $\g(\kp_0,\Del)$ for given $N_4$.
Our analysis assumes that the values of $\om(\kp_0,\Del)$ and
$\g(\kp_0,\Del)$ will only change little with increasing $N_4$. This assumption is well tested inside phase $C$,
and for the fixed four-volume we have been using, namely, $N_4^{(4,1)}\! =\! 40.000$.
Any significant changes in $\om$ and $\g$ must therefore be due to 
changes in the bare couplings $\kp_0$ and $\Del$.
A dense grid of measuring points in coupling constant space was used to collect the relevant data. 
Details of this computing-intensive process will be published elsewhere.
The resulting contour plots for  
$\om(\kp_0,\Del)$ and $\g(\kp_0,\Del)$ in the $(\kp_0,\Del)$-plane
(Fig.\ \ref{fig1}) can be interpreted directly in terms of constant physics: moving along any given line of
constant $\omega$ on the left contour plot, we can read off from the right contour plot how $\gamma$ 
changes along this line, and in particular whether it increases as desired for a UV limit.

\begin{figure}[t]
\vspace{-1cm}
\begin{center}
\includegraphics[width=0.65\textwidth]{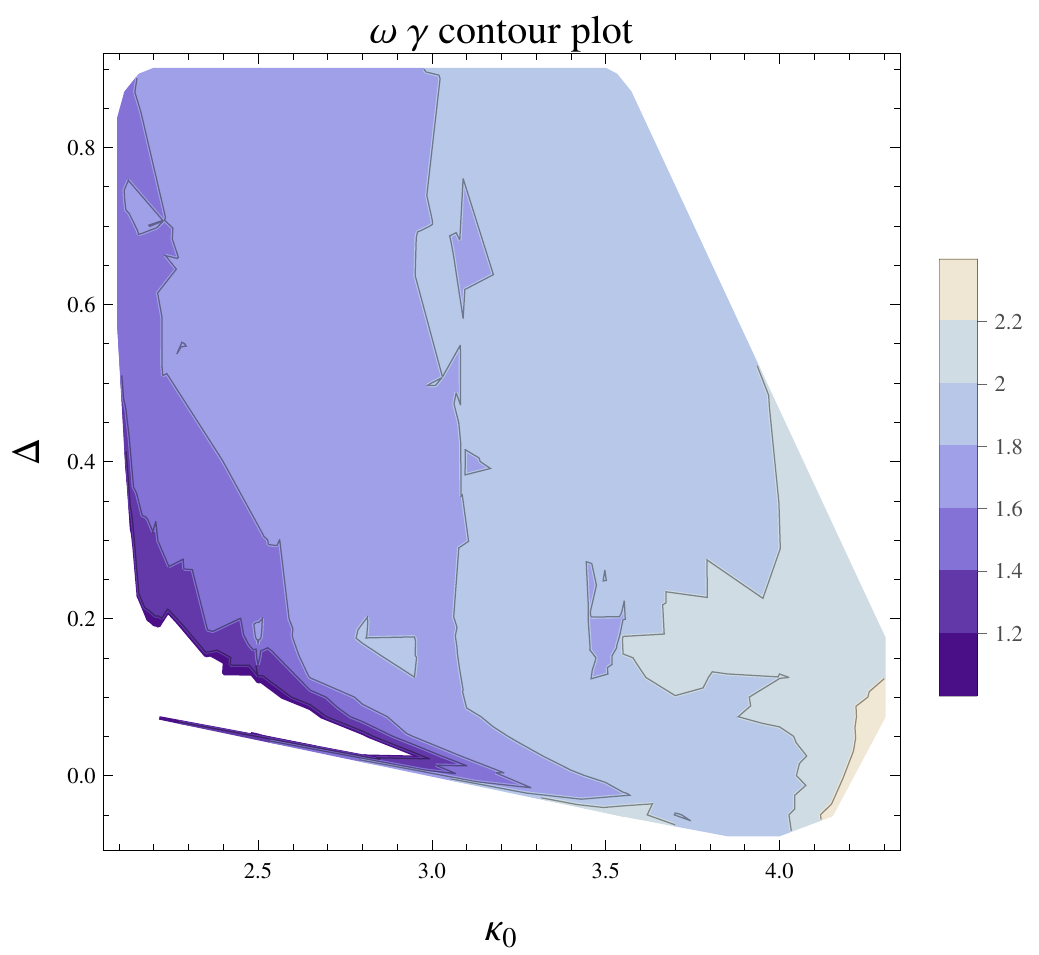}
\end{center}
\caption{Contour plot of the product $\om(\kp_0,\Del)\cdot\g(\kp_0,\Del)$ on 
coupling constant space.} 
\label{fig2}
\end{figure}

Approaching the $B$-$C$ phase boundary, which lies along the bottom of the two plots of Fig.\ \ref{fig1}, 
we observe that $\om$ decreases significantly, while $\g$ increases somewhat, as one
would expect when approaching a second-order phase transition
line. However, this increase does not appear to be large enough to result in an increase of
the product $\om(\kp_0,\Del)\cdot\g(\kp_0,\Del)$. According to the logic outlined above, this product 
should go to infinity in a UV limit where $V_4$ and $\kp$ 
stay constant while we take $a \to 0$. At least in the region where we can measure reliably, 
somewhat away from the transition line,
the product $\om\cdot \g$ changes little, as can be seen in Fig.\ \ref{fig2}.
Close to the $B$-$C$ phase transition line our results are
not reliable. Autocorrelation times grow enormously, and
the decrease in the parameter $\om$ means that the universe 
becomes very short in the time direction, rendering the use of 
the effective action \rf{2.4c} questionable.  

\section{UV fixed point scenario}

In the minisuperspace action \rf{2.4s} we have introduced a generalized 
inverse gravitational coupling constant $\kp$, which has 
mass dimension 2 and incorporates a dependence on the HLG-parameter $\lambda$. 
We can introduce a corresponding dimensionless coupling $\hat\kappa$ via 
$\kp(a)\! =\! \hat{\kp}(a)/a^2$. Comparing with relations \rf{2.4v}, 
we see that  $\hat{\kp}(a) \propto (\om/\chi^{3/4})^2 k_1(a)$. 
Of course, this identification is only meaningful as long as physics is well described
by the effective actions \rf{2.4s} and \rf{2.4c}. 
At least well inside phase $C$ this is known to be the case. 

For long-distance physics we expect $\kp$ to be a constant,
implying that $k_1$ should behave like $k_1(a) \propto \kp \cdot a^2 \propto \kp\, (V_4/N_4)^{1/2}$.
This implies the scaling behaviour $\g \propto 1/\sqrt{k_1} \propto N_4^{1/4}$, which 
we have already discussed earlier as a requirement of constant physics.
However, this is not the behaviour one would in general expect to encounter at a UV fixed point. 
By definition a nonperturbative UV fixed point is 
one where the {\it dimensionless} coupling goes to a finite fixed value,
$\hat{\kp}(a) \to \hat{\kp}^*$. Consequently, the analogue 
of the expansion \rf{2.2} for the inverse gravitational coupling constant is given by
\beq\label{2.8}
\hat{\kp}(N_4) = \hat{\kp}^* +const. \; N^{\bt'/4}_4,~~~~\bt' <0,
\eeq
provided we are in the vicinity of the fixed point $\hat{\kp}^*$ and 
move on a trajectory where $V_4$ is kept constant.
According to relations \rf{2.4v}  and \rf{2.8} this implies a $k_1$-behaviour of the form 
\beq\label{3.1}
k_1(N_4) \propto \left(\frac{\om}{\chi^{3/4}}\right)^2  \hat{\kp}^* ~~~({\rm large}\; N_4).
\eeq 
Still assuming that our minisuperspace analysis provides a reliable frame of reference, this leads to
\beq\label{3.2}
\frac{|\del V_3(\tau_i)|}{V_3(\tau_i)} = \frac{|\del N_3(i)|}{\langle N_3(i)\rangle } \propto 
\frac{\om(\kp_0(N_4),\Del(N_4))}{ \sqrt{k_1(N_4)} N_4^{1/4}} \propto 
\frac{\chi(\kp_0(N_4),\Del(N_4))^{3/4}}{\hat{\kp}^* N_4^{1/4}}~~~({\rm large}\; N_4).
\eeq
We conclude that this quotient cannot be kept constant in the 
neighbourhood of the UV fixed point and for constant $\chi$, unless 
for some reason $\hat{\kp}^*\! =\! 0$. One way to make a vanishing fixed-point value
for $k_1$ appear
natural is by explicitly invoking the HLG-parameter $\lambda$ and discussing 
the UV fixed point in terms of the coupling constant $\tkp$, which appears in the continuum action
\rf{2.4q}. In terms of its dimensionless counterpart $\hat{\tkp}(a)\! :=\! a^2 \tkp(a)$
one would make an ansatz
\beq\label{3.3} 
\hat{\tkp}(N_4) = \hat{\tkp}^* +const. \; N^{\bt'/4}_4,~~~\bt' <0,
\eeq
analogous to \rf{2.8}. However,
because of $\kp\! =\!  (\frac{1}{3}-\lam)\tkp$, in place of relation \rf{3.1} one then obtains 
\beq\label{3.4}
k_1(N_4) \propto \left(\frac{1}{3}-\lam\right) \left(\frac{\om}{\chi^{3/4}}\right)^2\,\hat{\tkp}^*~~~({\rm large}\; N_4).
\eeq 
This now leaves open the possibility of a vanishing $k_1$ at the ultraviolet fixed point, $k_1(N_4)\! \to\! 0$,
{\it provided} one chooses to scale $\lam \to 1/3$ at the same time. 
Note that by doing so one gives up staying 
on a curve of constant physics, in the sense of keeping $V_4$, $V_3$, $|\del V_3|$
and the shape of the emergent semiclassical minisuperspace geometry
fixed. The reason is that according to 
\beq\label{3.5}
\chi^2 = \frac{1-3\lam}{2\tilde{\del}}
\eeq
(c.f.\ eq.\ \rf{2.4s} and \rf{2.4u}) a change in $\lam$ implies a change in $\chi$, the parameter describing the 
shape of the universe, unless we choose to scale $\tilde{\del}$ precisely as $\tilde{\del} \!\propto\! (1-3\lam)$.
According to our assumptions, $\om$ then also changes. 
If $\tilde{\del}$ stays constant or goes to zero slower than $(1-3\lam)$, both
$\chi\to 0$ and $\om \to 0$ at the UV fixed point. Since 
we observe in our computer simulations that $\om$ goes towards zero when we approach the 
$B$-$C$ second-order phase transition line, the line
appears as a candidate for UV fixed points in this particular scenario. Approaching it
along some path where $\chi$ (and therefore $\om$) decreases
but $V_4$ is kept fixed implies that $V_3 \propto  (V_4/\chi)^{3/4}$
is no longer constant. Also the constancy criterion \rf{2.6} for
$|\del V_3|/V_3$ can no longer be applied in a straightforward manner.

If on the other hand we choose to scale $\tilde{\del}$ like $(1-3\lam)$, we can maintain
the concept of constant shape and three-volume $V_3$ for fixed $V_4$. 
We are then back to the situation analyzed previously; $\g$ has to 
grow proportional to $N_4^{1/4}$ along paths of constant $\om$, with the 
only difference that this now allows for a UV interpretation
in terms of $\tkp$ rather than $\kp$. However, as discussed in the previous section, 
there is little support for this growth from the data, at least in the region where we can
measure reliably.

\section{Discussion and conclusion}\label{uvlimit}

In this paper, we have presented the results of a first nonperturbative
analysis of renormalization group flows in four-dimensional CDT quantum gravity.
Since a second-order phase transition line has been found in this formulation of quantum
gravity \cite{phase1} -- thus far a very rare occurrence in dynamical models of higher-dimensional
geometry -- how this line may be reached along suitably defined RG trajectories in phase
space will give us important information about the theory's ultraviolet regime. It will also
allow us to make a closer comparison with continuum investigations of gravity in terms
of functional renormalization group techniques and may provide an independent check
on ultraviolet fixed point scenarios derived in this approach.

As explained in Secs.\ \ref{sec:constant} and \ref{sec:apply}, we use conventional lattice methods to investigate
the behaviour near the phase transition, adapted to the case of dynamical geometry, where we
do not have a fixed background geometry to refer to and any physical yardstick for measuring
distances has to be generated dynamically. Taking a UV limit is achieved formally by
sending the lattice spacing $a$ to zero, but to make this into a physically meaningful prescription
$a$ has to be related to some {\it physical} length units. As illustrated by the scalar field example, 
this is usually done by referring to the correlation length. Alternatively, since in the case of
gravity we currently do not have a suitable correlation length available, one may also refer to the
total volume of the system, and re-express scaling relations near a fixed point in terms of this volume,
as illustrated by eq.\ \rf{2.2}. This is the strategy we follow for gravity to make sure that we have
a true, physical implementation of the ultraviolet limit. The difference with the scalar field case is
that the macroscopic reference volume used is generated {\it dynamically}, and any possible 
dependence on the bare couplings should be considered carefully, because it can have
an influence on how one defines `lines of constant physics' on 
coupling constant space.

For the latter we have made the most direct ansatz available in CDT gravity, namely, to define
constant physics in terms of the physical quantities characterizing the macroscopic universe
that emerges as the ground state of the quantum dynamics. These are its total four-volume, its 
three-volume as a function of proper time and quantum fluctuations of the three-volume around
its mean, the so-called volume profile. We have interpreted all of them physically in terms of
a class of homogeneous and isotropic cosmological solutions of Ho\v rava-Lifshitz 
type, and have assumed that this interpretation is valid throughout phase $C$, where we observe 
extended geometry. At the same time we have assumed that we can make an identification of
lattice units in terms of continuum proper times and distances that likewise remains unchanged
inside phase $C$. Conceptually, these are the most straightforward assumptions one can make,
and it is important to understand what conclusions they lead to. 

Concretely, we then defined lines of constant physics by keeping the shape parameter $\om(\kp_0,\Del)$ constant,
as well as the relative size of three-volume fluctuations, leading to the scaling requirement
$\g(\kp_0,\Del) \propto N_4^{1/4}$ for the ``fluctuation parameter" $\gamma$ in the UV limit
$N_4\to\infty$.
Analyzing the computer simulation data presented above we saw no concrete indication
that the second-order $B$-$C$ phase transition line is reached when flowing along any of the
lines of constant physics. Instead, the lines of constant $\om(\kp_0,\Del)$ run parallel to 
the B-C phase transition line if one starts in the vicinity of this line. 
Increasing $\g(\kp_0,\Del)$ along such a line brings
one close to the triple point of the phase diagram. For the 
finite value of $N_4$ used here, curves of constant $\om$ 
eventually turn away from the triple point and run parallel to 
the $A$-$C$ transition line. However, this may well be a finite-volume effect,
leaving open the possibility for flow lines to end up in the triple point.
On the other hand, on the basis of the measurements made up to now the increase 
in $\g$ when moving along a line of constant $\om$ seems to 
be too slow to satisfy the criteria of constant physics for $N_4 \to \infty$.

However, as we described in Sec.\ \ref{uvlimit}, it {\it is} possible to view the $B$-$C$ line as a 
second-order UV phase 
transition line for the HLG action \rf{2.4q}, if we allow for a suitable scaling of ``little lambda", 
$\lam \to 1/3$.\footnote{The value $\lam\! =\! 1/3$ is special in the sense that the Wheeler-DeWitt 
metric underlying
the construction of the kinetic term in the action \rf{2.4q} becomes degenerate, and
an extra constraint appears in the Hamiltonian analysis. HLG models setting
$\lam\! =\! 1/3$ from the outset have recently been considered in their own right \cite{brs}.}
In this interpretation an anisotropy between 
space and time develops as one moves along flow lines, corresponding to
$\chi \to 0$ in \rf{2.4x}. 

It is clear that the next step in our investigation of renormalization group flows 
will be a more extended analysis of different UV
scaling scenarios, where in particular our current assumption of ``frozen" proper distance units
throughout coupling constant space is relaxed, which will have consequences for how ``constant physics"
is defined. It would also allow us to consider a scenario where the shape of the emergent
universes is interpreted in terms of round four-spheres, at least somewhat away from the
phase transition, as we have done in previous work \cite{ajl,agjl}, in contrast to the family of
deformed spheres we have used here. It is clear that this can change the running of the 
renormalization group flows significantly, and improve on the results found in the present
work, where we have adopted rather conservative assumptions about scaling and constant
physics.

Using different notions of constant physics close to the phase transition is certainly well motivated
by nonclassical features of quantum geometry already found on Planckian scales, like the anomalous
behaviour of the spectral dimension \cite{spectral}, and by taking seriously anisotropic scaling scenarios 
\`a la Ho\v rava in the UV, which we have already argued constitute a natural frame of reference for our
investigation. There will be technical issues to deal with when investigating different scalings
near the $B$-$C$ transition line, including the fact that the time extension of the universe
shrinks to only a few lattice spacings there, making any construction
of an effective action imprecise. One obvious solution would be to increase the lattice size $N_4$,
but one also has to take into account the critical slowing-down near the $B$-$C$ transition
(as one would expect), which makes simulations there painfully slow. 
We are currently trying to circumvent this issue by using 
the so-called transfer matrix formalism \cite{transfer}, 
where a large time extension is not needed. Progress on this will be reported elsewhere.

\vspace{.5cm}      
\noindent {\bf Acknowledgments.} 
JA and AG  acknowledge support from the ERC Advanced Grant 291092
``Exploring the Quantum Universe'' (EQU) and by FNU, 
the Free Danish Research Council, through the grant 
``Quantum Gravity and the Role of Black Holes''. JJ acknowledges the 
support of grant DEC-2012/06/A/ST2/00389 from 
the National Science Centre Poland. 
The contributions of AK and RL are part of the research programme of the Foundation for Fundamental Research 
on Matter (FOM), financially supported by the Netherlands Organisation for Scientific Research (NWO).
The work was also sponsored by NWO Exacte Wetenschappen (Physical Sciences) for the use 
of supercomputer facilities, with financial support from NWO. 
JA and RL were supported in part by Perimeter Institute of Theoretical Physics.
Research at Perimeter Institute is supported by the Government of Canada
through Industry Canada and by the Province of Ontario through the 
Ministry of Economic Development \& Innovation. The authors thank 
Daniel Coumbe for discussion, for reading the 
paper and for useful comments.


\begin{thebibliography}{99}

\bibitem{physrep}
J.~Ambj\o rn, A.~G\"orlich, J.~Jurkiewicz and R.~Loll,
Phys.\ Rept.\  {\bf 519} (2012) 127 [arXiv:1203.3591, hep-th].

\bibitem{cdtreviews}
J.~Ambj\o rn, J.~Jurkiewicz and R.~Loll,
in {\it Approaches to Quantum Gravity}, ed. D.\ Oriti (Cambridge
University Press, Cambridge, UK, 2009) 341-359
[hep-th/0604212].\\
J.~Ambj\o rn, A.~G\"orlich, J.~Jurkiewicz and R.~Loll,
in {\it Path Integrals - New Trends and Perspectives}, eds. W.\ Janke
and A.\ Pelster (World Scientific, Singapore, 2008)
191-198;
Acta Phys.\ Polon.\ B\ {\bf 39} (2008) 3309.\\
J.~Ambj\o rn, J.~Jurkiewicz and R.~Loll,
Annalen Phys.\ {\bf 19} (2010) 186;
in {\it Foundations of Space and Time}, eds. G.\ Ellis, J.\ Murugan and A.\ Weltman
(Cambridge University Press, Cambridge, UK, 2012)
[arXiv: 1004.0352, hep-th];
PoS LATTICE {2010} (2010) {\bf 014}
[arXiv: 1105.5582, hep-lat].

\bibitem{weinberg}
{S.~Weinberg},
in {\sl General Relativity: Einstein Centenary Survey}, 
eds. S.W.\ Hawking and W.\ Israel
(Cambridge University Press, Cambridge, UK, 1979) 790-831.

\bibitem{kawai}
  { H.~Kawai and M.~Ninomiya,}
  Nucl.\ Phys.\ B {\bf 336} (1990)  115.\\
  { H.~Kawai, Y.~Kitazawa and M.~Ninomiya,}
  Nucl.\ Phys.\ B {\bf 393} (1993)  280-300
  [hep-th/9206081];
  Nucl.\ Phys.\ B {\bf 404} (1993)  684-716
  [hep-th/9303123];
Nucl.\ Phys.\ B\ {\bf 467} (1996) 313-331 [hep-th/9511217].\\
  { T.~Aida, Y.~Kitazawa, H.~Kawai and M.~Ninomiya,}
  Nucl.\ Phys.\ B {\bf 427} (1994)  158-180
  [hep-th/9404171].

\bibitem{FRG}
  M.~Reuter,
  Phys.\ Rev.\ D {\bf 57} (1998)  971-985
  [hep-th/9605030].\\
A.~Codello, R.~Percacci and C.~Rahmede,
Annals Phys.\  {\bf 324} (2009) 414 [arXiv:0805.2909, hep-th].\\
M.~Reuter and F.~Saueressig,
[arXiv:0708.1317, hep-th].\\
M.~Niedermaier and M.~Reuter,
Living Rev.\ Rel.\ {\bf 9} (2006) 5.\\
D.F.~Litim,
Phys.\ Rev.\ Lett.\ {\bf 92} (2004) 201301 [hep-th/0312114].

\bibitem{horava}
  P.~Ho\v rava,
  Phys.\ Rev.\ D {\bf 79 } (2009)  084008
  [arXiv:0901.3775, hep-th].\\
  P.~Ho\v rava and C.M.~Melby-Thompson,
  Phys.\ Rev.\ D {\bf 82} (2010)  064027
  [arXiv:1007.2410, hep-th].

\bibitem{spectral} 
  J.~Ambj\o rn, J.~Jurkiewicz and R.~Loll,
  Phys.\ Rev.\ Lett.\  {\bf 95} (2005) 171301 [hep-th/0505113].\\
  O.~Lauscher and M.~Reuter,
  JHEP {\bf 0510} (2005) 050
  [hep-th/0508202].\\
  P.~Ho\v rava,
  Phys.\ Rev.\ Lett.\  {\bf 102} (2009) 161301
  [arXiv:0902.3657, hep-th].\\
  M.~Reuter and F.~Saueressig,
  New J.\ Phys.\  {\bf 14} (2012) 055022
  [arXiv:1202.2274, hep-th].

\bibitem{phase}
J.~Ambj\o rn, A.~G\"orlich, S.~Jordan, J.~Jurkiewicz and R.~Loll,
  Phys.\ Lett.\ B {\bf 690} (2010) 413
  [arXiv:1002.3298, hep-th].\\
P.~Ho\v rava,
  Class.\ Quant.\ Grav.\  {\bf 28} (2011) 114012 [arXiv:1101.1081, hep-th].

\bibitem{phase1}
  J.~Ambj\o rn, S.~Jordan, J.~Jurkiewicz and R.~Loll,
  Phys.\ Rev.\ Lett.\  {\bf 107} (2011) 211303
  [arXiv:1108.3932, hep-th];
  Phys.\ Rev.\ D {\bf 85} (2012) 124044
  [arXiv:1205.1229, hep-th].

\bibitem{GFT}
D.~Oriti,
in {\it Approaches to Quantum Gravity}, ed. D.\ Oriti (Cambridge
University Press, Cambridge, UK, 2009) 310-331
[gr-qc/0607032];
Class.\ Quant.\ Grav.\  {\bf 27} (2010) 145017 [arXiv:0902.3903, gr-qc].

\bibitem{tensor}
  V.~Bonzom, R.~Gurau, A.~Riello and V.~Rivasseau,
  Nucl.\ Phys.\ B {\bf 853} (2011) 174
  [arXiv:1105.3122, hep-th].\\
  R.~Gurau and J.P.~Ryan,
  SIGMA {\bf 8} (2012) 020
  [arXiv:1109.4812, hep-th].

\bibitem{aj}
J. Ambj\o rn, J. Jurkiewicz, Phys. Lett. {\bf B} 278 (1992) 42.
 
\bibitem{am}
  M.E.~Agishtein and A.A.~Migdal,
Mod.\ Phys.\ Lett.\ A {\bf 7} (1992) 1039;
  Nucl.\ Phys.\ B {\bf 385} (1992) 395.

\bibitem{hl-rg}
  S.~Rechenberger and F.~Saueressig,
  JHEP {\bf 1303} (2013) 010
  [arXiv:1212.5114, hep-th].

\bibitem{crs}
A.~Contillo, S.~Rechenberger and F.~Saueressig,
  JHEP {\bf 1312} (2013) 017 [arXiv:1309.7273, hep-th].

\bibitem{carlip}
C.~Anderson, S.J.~Carlip, J.H.~Cooperman, P.~Ho\v rava, 
R.K.~Kommu and P.R.~Zulkowski,
  Phys.\ Rev.\ D {\bf 85} (2012) 044027
  [arXiv:1111.6634, hep-th].
  
\bibitem{jordanloll}
  S.~Jordan and R.~Loll,
  Phys.\ Lett.\ B\ {\bf 724} (2013) 155-159 [arXiv:1305.4582, hep-th];
  Phys.\ Rev.\ D {\bf 88} (2013) 044055 [arXiv:1307.5469, hep-th].

\bibitem{original}
  J.~Ambj\o rn, J.~Jurkiewicz and R.~Loll,
  Nucl.\ Phys.\ B {\bf 610} (2001) 347
  [hep-th/0105267].

\bibitem{ajl}
  J.~Ambj\o rn, J.~Jurkiewicz and R.~Loll,
  Phys.\ Rev.\ D {\bf 72} (2005) 064014
  [hep-th/0505154].

\bibitem{ajl1}
J.~Ambj\o rn, J.~Jurkiewicz and R.~Loll,
Phys.\ Rev.\ Lett.\  {\bf 93} (2004) 131301 [hep-th/0404156].

\bibitem{agjl}
  J.~Ambj\o rn, A.~G\"orlich, J.~Jurkiewicz and R.~Loll,
  Phys.\ Rev.\ D {\bf 78} (2008) 063544
  [arXiv:0807.4481, hep-th];
Phys.\ Rev.\ Lett.\  {\bf 100} (2008) 091304 [arXiv:0712.2485, hep-th].

\bibitem{agjl-etal}
J.~Ambj\o rn, A.~G\"orlich, J.~Jurkiewicz, R.~Loll, 
J.~Gizbert-Studnicki and T.~Trzesniewski,
Nucl.\ Phys.\ B {\bf 849} (2011) 144 [arXiv:1102.3929, hep-th].

\bibitem{Ambjorn:2004pw}
J.~Ambj\o rn, J.~Jurkiewicz and R.~Loll,
 Phys.\ Lett.\ B {\bf 607} (2005) 205 [hep-th/0411152].

\bibitem{brs}
J.~Bellorin, A.~Restuccia and A.~Sotomayor,
Phys.\ Rev.\ D {\bf 87} (2013) 8,  084020 [arXiv:1302.1357, hep-th].

\bibitem{transfer}
  J.~Ambj\o rn, J.~Gizbert-Studnicki, A.~G\"orlich and J.~Jurkiewicz,
  JHEP {\bf 1209} (2012) 017
  [arXiv:1205.3791, hep-th].


\end{thebibliography}
\end{document}